\documentclass[english,twocolumn,showpacs,amsmath,amssymb,superscriptaddress]{revtex4}
\usepackage{graphicx}
\usepackage{amsmath}
\begin{document}

\title{\bf NeXSPheRIO results on azimuthal anisotropy in Au-Au collisions at 200A GeV}

\author{Wei-Liang Qian}
\author{Rone Andrade}
\author{Arthur dos Reis}
\author{Yogiro Hama}
\author{Fr\'{e}d\'{e}rique Grassi}

\affiliation{Instituto de F\'{\i}sica, USP, C. P. 66318, 05314-970 S\~{a}o Paulo, SP, Brazil}

\author{Takeshi Kodama}

\affiliation{Instituto de F\'{\i}sica, UFRJ, C. P. 68528, 21945-970 Rio de Janeiro-RJ , Brazil}

\begin{abstract}
In this work, we present the results obtained by the hydrodynamic code NeXSPheRIO on anisotropic flows. 
In our calculation, we made use of event-by-event fluctuating 
initial conditions, and chemical freeze-out was explicitly implemented.
We studied directed flow, elliptic flow and forth harmonic coefficient 
for various hadrons at different centrality windows for Au+Au collisions at 200 AGeV. 
The results are discussed and compared with experimental data from RHIC. 
\end{abstract}
\vspace{12pt}
\pacs{25.75.-q, 24.10.Nz, 25.75.Ld}

\maketitle

\section{Introduction}
Anisotropic flows are defined by the Fourier decomposition of the azimuthal distribution of particles produced in 
heavy ion collisions with respect to the event plane. 
Directed flow and elliptic flow are quantified by the first harmonic ($v_1$) and the second harmonic ($v_2$), respectively.
The study of anisotropic flows in high energy nuclear collisions provides us valuable information on equation of state (EOS), degree of thermalization and the early stage of the created hot, dense matter. It also plays an important role in the extensively discussed question whether the quark-gluon plasma (QGP) created in Au+Au collisions at the Relativistic Heavy Ion Collider (RHIC) represents a "perfect" liquid. 

Hydrodynamics offers a reasonable description of the experimental data for bulk properties of low $p_T$ particles for heavy-ion collisions 
at RHIC\cite{hydro1,hydro2,hydro3,hydro4,hydro5,topics}. 
Besides, the discovery of the "perfect" liquid\cite{pf-exp,pf-th} at RHIC in Brookhaven National Laboratory (BNL)
was based on the successful modelling of the anisotropic expansion of the matter in the early 
stage of the reaction by means of ideal hydrodynamics. 
Specifically, at the RHIC energy of 200 A GeV, 
the observed $v_2$ values agree well with predictions from ideal fluid dynamics\cite{v2hydro} 
in the midrapidity region ($|\eta| \lesssim 1$) with small or intermediate impact parameters and transverse momenta.
The agreement between model predictions and the experimental data suggests that hydrodynamic model is 
well justified to describe the intermediate stage of the reaction. 
It is the hydrodynamics that plays an important role 
to fill the gap between the static aspects of QGP properties 
and dynamical aspects of heavy ion collisions measured experimentally.

In this work, we present recent results on azimuthal anisotropy obtained by a hydrodynamic code developed by our group, 
called NeXSPheRIO. The code is based on full three dimentional ideal hydrodynamics. We calculated 
directed flow coefficient $v_1$, elliptic flow coefficent $v_2$ and forth harmonic coefficient $v_4$ 
of Au+Au collisions at 200 A GeV at RHIC, 
for various identified particles at different centrality windows. 
Two important aspects are employed in our present approach, namely, fluctuating initial conditions (IC) and chemical freeze-out.
Their effects on observables are addressed. 

In what follows, we will first give a brief description of the NeXSPheRIO code and 
how this code is used to compute the observables of our interest. 
The results are shown in the third section, along with discussions and perspectives. 

\section{The NeXSPheRIO Code}

The numerical tool for our present study is called NeXSPheRIO. 
It is a conjunction of two codes, NeXus and SPheRIO. 
NeXus\cite{nexus} is as an event generator.
It is a microscopic model based on Gribov-Regge theory\cite{regge},
which provides the initial conditions (IC) for a macroscopic treatment of the hydrodynamic stage of the collision.
The IC state is given as spacial distributions of energy momentum tensor and other conserved charge currents. 
We assume, as in most hydrodynamic models on the market, that the local equilibrium is reached within a short time scale $\tau \sim$ 1 fm/$c$. 
Owing to the partonic structure of high momentum incident nuclei, the IC obtained at a very early instant can not be smooth, 
but highly inhomogeneous in the space. 
In addition, since the size of the system we are dealing with is very small, 
big event-by-event fluctuations are expected in real collisions. 
This feature is very different from most hydrodynamic treatements where averaged smooth IC are usually adopted.
As already observed and studied previously in \cite{ic1,ic2,ic3,ic4,ic4a5,ic5}, such fluctuations in IC 
lead to significant effects on experimental observables. 
In this context, one of the advantages of NeXus is that it can produce realistic inhomogeneous IC in event-by-event basis, 
therefore provides us with a good opportunity to study the effect of fluctuations in IC. We show in Fig.1 an example of one such fluctuating
event, generated by NeXus, for a random central Au+Au collision at 200AGeV. As one can see, the energy-density
distribution presents several blobs where energy is highly concentrated. 

To solve the hydrodynamic evolution for systems like the one shown in Fig.1 is not a trivial task. 
The section of our code which undertakes this job is called SPheRIO\cite{topics}. The code is based on the 
Smoothed Particle Hydrodynamic(SPH) algorithm, 
which was first introduced for astrophysical applications\cite{astro}, and later adopted in relativistic heavy-ion collisions\cite{va}. 
The model parameterizes the matter flow in terms of discrete Lagrangian coordinates, called SPH particles. 
In terms of SPH degrees of freedom, the equations of motion can be derived by using the variational principle\cite{vp}. 
The main advantage of the method is that it is rather robust to deal with any kind of geometrical structure and violent dynamics. 
At the meantime, it is reliable and efficient, provided the size of SPH particles is appropriately chosen.

We use in the present study an equation of state (EOS) which takes into consideration 
the strangeness conservation. 
We adopt hadronic resonance model with finite volume corrections to describe the matter in the hadronic side, 
where the main part of observed resonances in Particle Data Tables\cite{pdtable} has been included. 
For quark gluon plasma phase, the MIT bag model is made use of. 
By solving the Gibbs equation, one obtains a first order phase transition in the mixed phase. 
As a good approximation, local strangeness neutrality is assumed. We shall neglect any dissipative effects in this work. 

As it was shown\cite{cfo-qian}, hydrodynamic model without chemical freeze-out gives good
description of transverse momentum spectra of light hadrons.
While it reproduces the shape of spectra for hyperons and anti-protons, 
there are visible discrepancies in the multiplicities. 
To compensate this, we implemented early chemical freeze-out 
for strange hadrons such as $\Lambda$, $\Xi$, $\Omega$ and $\phi$ since their cross sections are small.
At chemical freeze-out temperature $T_{ch}$, 
these particles cease to have inelastic collisions and therefore their abundances are fixed. 
We assume complete chemical equilibrium on the surface of chemical freeze-out, 
thus no strangeness saturation factor ${\gamma}_{S}$ is introduced in our treatment. 
In our calculation, $T_{ch}$ and $T_{th}$ serve as adjustable parameters
as function of centrality according to experimental data. 
The introduction of chemical freeze-out for hyperons into hydrodynamics model helps to correctly reproduce the hyperon yields 
of transverse momentum spectra. 
In the present work, above mentioned chemical freeze-out is employed.

\section{Results and discussions}

Here, we first tune the free parameters of the model to match the experimental data for the 
multiplicity and radial flow. 
We begin by fixing the IC so as to reproduce properly the pseudo-rapidity distributions of charged particles in each centrality window. 
This is done by applying an $\eta$-dependent factor $\sim{1}$ \cite{ic4a5} to the initial energy-density distributions for each centrality window. 
Next, to correctly reproduce the transverse momentum spectra of charged particles, 
we appropriately parametrize the centrality dependent freeze-out temperature $T_{th}$. 
As for chemical freeze-out temperatures, they are adjusted to reproduce the yield of $\Lambda$s. 
We calculate the anisotropic flows of identified particles with respect to the event planes of all charged particles. 
As in our previous studies\cite{ic4,ic4a5,ic5}, the event planes are defined using PHOBO subevent method\cite{subevent}.

Our main results are shown in Fig.2 - Fig.6. We plot in Fig.2 - Fig.4 the elliptic flow coefficient $v_2$ of pions, protons, anti-protons and $\Lambda$s 
for diferent centrality windows and minimal bias. In Fig.5 and Fig.6, directed flow coefficient $v_1$ and the fourth harmonic coefficient $v_4$ 
for all charged particles are depicted. 
The numerical results are compared with data from STAR Collaboration\cite{star4}. 
One finds that $v_2$ and $v_4$ increase with increasing transverse momentum as well as with more peripheral collisions, 
and $v_2$ decreases when the mass of identified particle increases. 
It is observed that the results obtained by using sudden Cooper-Frye freeze-out undershoot the $v_2$ data 
of pions and slightly overestimate those of protons, anti-protons and $\Lambda$s 
in small transverse momentum region with $p_T \leq 2$,
but overall they are in qualitative agreement with the experimental data.
The result of directed flow coefficient $v_1$ shows reasonable agreement in small $\eta$ region. 
Since STAR data of $v_1$
do not extend to the regions in which $|\eta| > 4$, we are not able to compre the data there where the calculated curve turns over. 
Our result of $v_4$ stays little below the experimental data, but it demonstrates the correct trend.

A meaningful solution to improve the elliptic flow result is to adopt the continuous emission (CE) prescription\cite{ce1,ce2}.
In the CE picture, the emission of hadrons occurs not only from the sharply defined freeze-out hypersurface, 
but continuously from the whole expanding volume of the system at different temperatures and different times. 
As a result, the large-$p_T$ particles are mainly emitted at early instants when 
the fluid is hot and mostly from its surface.
One might imagine that isotropic emission of large $p_T$ components 
at the surface will naturally decrease $v_2$ at large transverse momentum region,
and the $p_T$ spectra will become more concave. 
These effects are very similar to those discussed in other recent studies where one incorporates viscosity explicitly into 
the equation of motion\cite{vishydro1,vishydro2,vishydro3,vishydro4},
or connects the hydrodynamic model to a hadronic transport model\cite{iccgc}.
On the other hand, small-$p_T$ components are mostly emitted at later times 
when the fluid is cooler and from larger spatial domain, which might lead to an enhancement of $v_2$ in small $p_T$ region.
For particles like $\phi$, it is usually understood that they have smaller cross sections\cite{phic},
so it is reasonable to treat their freeze-out process differently.
Consequently, the value of $v_2$ on light hadrons such as pions might increase thus the result will be improved,
and meanwhile those of heavy hyperons will remain unchanged.
Further work to improve and extend the present study is under progress. 

We acknowledge financial support by FAPESP (2004/10619-9, 2005/54595-9, 2008/55088-1), CAPES/PrOBRAL, CNPq, FAPERJ and PRONEX.

\begin{figure}[!htb]
\vspace*{0.5cm}
\includegraphics[width=8.5cm]{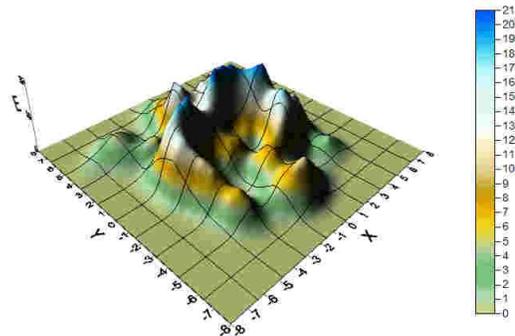}
\vspace*{-4cm}
\label{figure:fig1}
\caption{Energy density distribution of one random event at mid-rapidity plane for central 200 A GeV Au+Au collisions generated by NeXus. The diagram is plotted in units of GeV/fm$^3$. } 
\end{figure}

\begin{figure}[!htb]
\vspace*{-1.5cm}
\includegraphics[width=8.5cm]{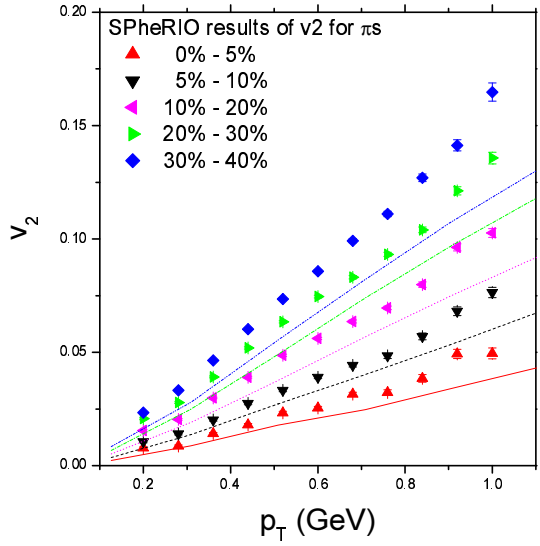}
\vspace*{-3.5cm}
\label{figure:fig2}
\caption{$v_2$ of pions as a function of $p_T$ at different centrality windows; the experimental data are from STAR Collaboration \cite{star4}.} 
\end{figure}

\begin{figure}[!htb]
\vspace*{-1.5cm}
\includegraphics[width=8.5cm]{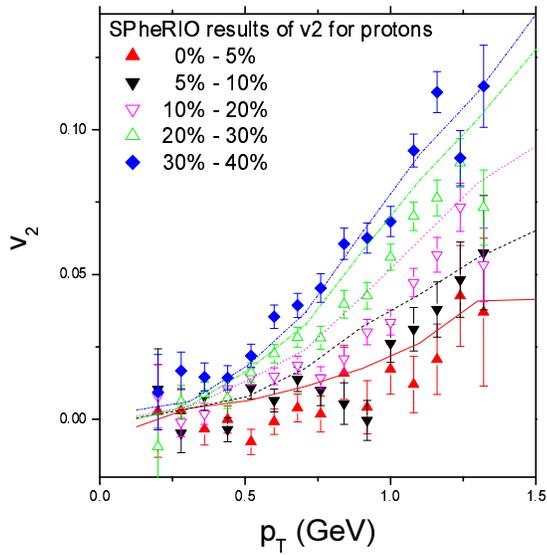}
\vspace*{-3.5cm}
\label{figure:fig3}
\caption{$v_2$ of protons and anti-protons as a function of $p_T$ at different centrality windows; the experimental data are from STAR Collaboration \cite{star4}.} 
\end{figure}

\begin{figure}[!htb]
\vspace*{-1.5cm}
\includegraphics[width=8.5cm]{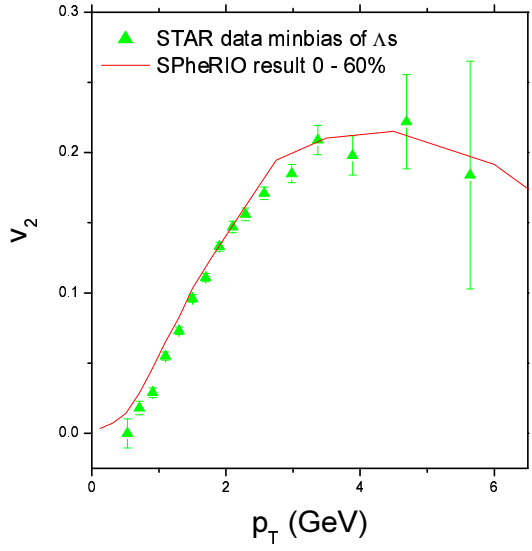}
\vspace*{-3.5cm}
\label{figure:fig4}
\caption{$v_2$ of $\Lambda$s as a function of $p_T$  at the centrality window as indicated; the experimental data are from STAR Collaboration \cite{star4}.} 
\end{figure}

\begin{figure}[!htb]
\vspace*{-1.5cm}
\includegraphics[width=8.5cm]{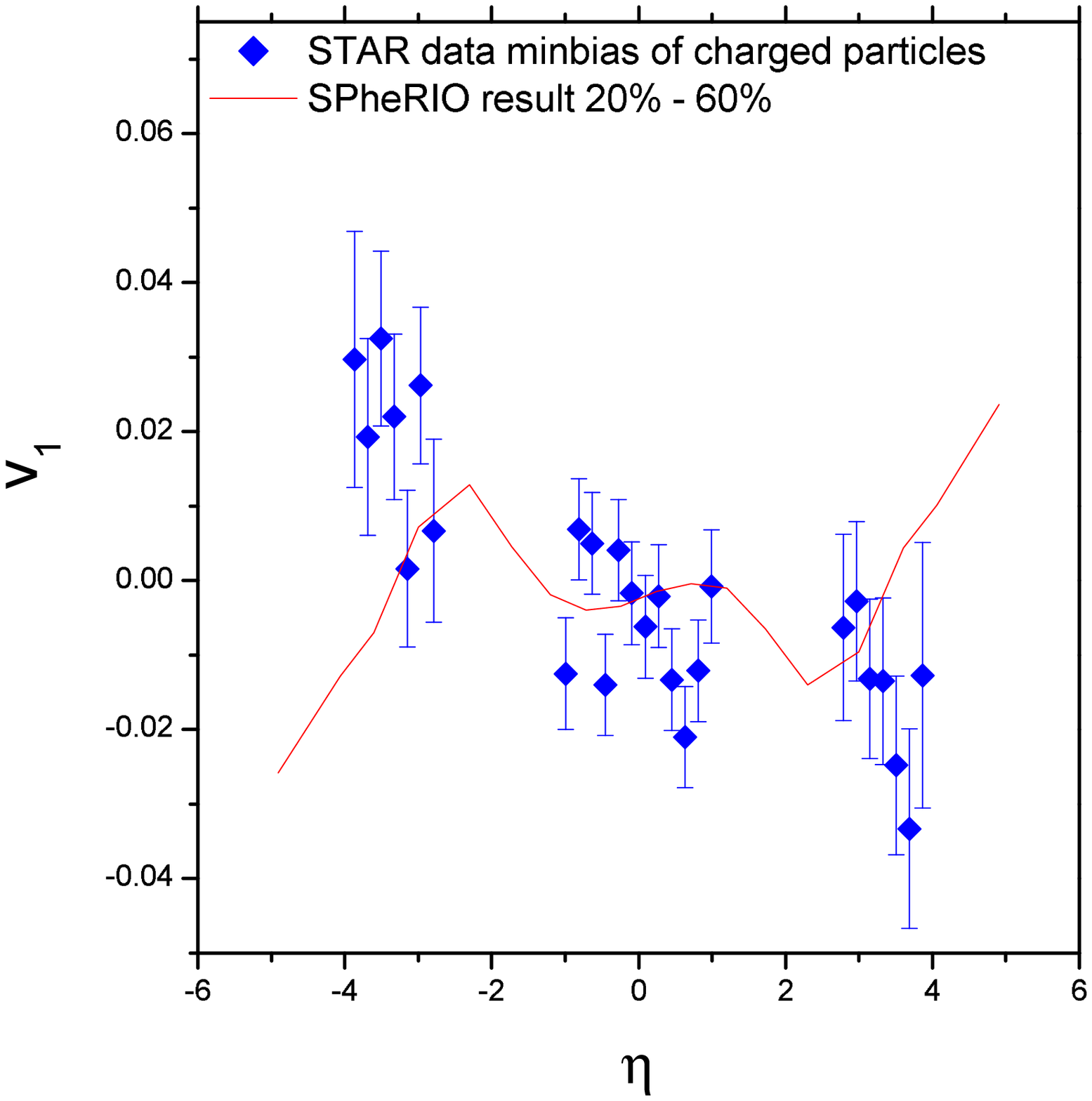}
\vspace*{-3.5cm}
\label{figure:fig5}
\caption{$v_1$ of all charged particles as a function of $p_T$; the experimental data are from STAR Collaboration \cite{star4}.} 
\end{figure}

\begin{figure}[!htb]
\vspace*{-1.5cm}
\includegraphics[width=8.5cm]{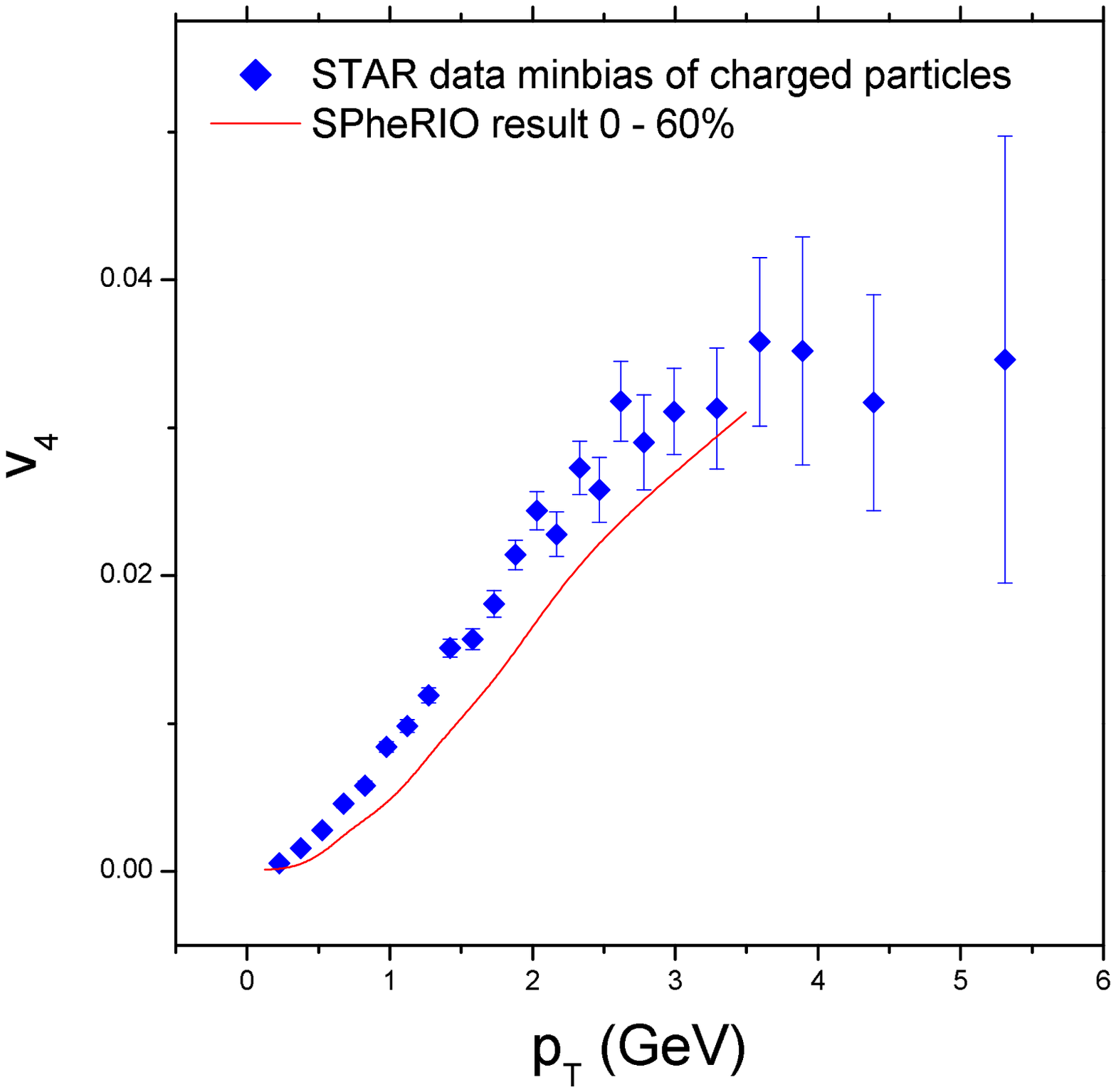}
\vspace*{-3.5cm}
\label{figure:fig6}
\caption{$v_4$ of all charged particles as a function of $p_T$; the experimental data are from STAR Collaboration \cite{star4}.} 
\end{figure}

\end{document}